\documentclass[10pt]{article}
\usepackage{bm}
\usepackage{psfrag}
\usepackage[dvips]{graphicx}

\newcommand{\ket}[1]{\ensuremath{\left|{#1}\right>}}
\newcommand{\even}{\ket{\mathrm{even}}}
\newcommand{\odd}{\ket{\mathrm{odd}}}

\begin{document}
\psfragscanon

\title{Manifestation of additional dimensions of space-time in mesoscopic systems}

\author{A.F.Andreev}
%\date{\today}
\date{}
\maketitle

\begin{center}
\small
\textit{P.L.Kapitza Institute for Physical Problems,
Russian Academy of Sciences,\\
Kosygin str., 2, 119334 Moscow, Russia}
\end{center}

\begin{abstract}
We note that the existence of physical states which are
coherent superpositions of
states with even and odd numbers of fermions means the
existence, together with $x,y,z,t$, of additional spinor
dimensions of space-time. A system with variable number of electrons is
described in which such superpositions are realized. Experiments with
mesoscopic condensed matter systems are suggested which generalize the
experiment of Nakamura et al. and may provide direct observation of
such superpositions and, thereby, justify the reality of a superspace with
additional spinor dimensions introduced in quantum field theory to account for
supersymmetry. The nature of additional dimensions of space-time is elucidated
for nonrelativistic systems.
\end{abstract}

\section{Introduction}
In 70-th the notion of superspace is introduced in quantum field
theory to make supersymmetry possible. In addition to the usual
dimensions (coordinates) $x,y,z,t$, spinor fermionic
(anticommuting) dimensions of space-time are introduced. If
supersymmetry is indeed discovered in the next generation of
experiments at Fermilab and CERN within a decade or two (see the review
paper \cite{ss}), this will amount to the discovery of the new dimensions.

There is another general problem in which the existence of
additional spinor coordinates of space-time plays an important
role. In 1952, Wick, Wightman, and Wigner \cite{www} show that the
coherent superpositions of states with even and odd numbers of
fermions are incompatible with the Lorenz invariance and introduce
the superselection rule, according to which such superpositions
are physically impossible. In actuality (as is pointed out in
\cite{a1,a2}), the superselection rule is the alternative to the
existence of additional spinor coordinates.

The point is that the vector \odd\ of any state
with an odd number of fermions, being spinor of the odd rank, is
multiplied by $(-1)$ upon the rotations $O(2\pi)$ of the
coordinate system through an angle of $2\pi$ about any axis and
upon the double time reversal $R^2$. The vectors \even\ 
of the states with an even number of fermions do not
change under the $O(2\pi)$ and $R^2$ transformations. For this
reason, the existence of a coherent superposition

\begin{equation}
\left| * \right> = u \even + v \odd
\label{eq1}
\end{equation}
where $u$ and $v$ are nonzero complex numbers, $|u|^2 + |v|^2 =
1$, implies the existence of a strange state $\left| * \right>$
which changes physically under the $O(2\pi)$ and $R^2$
transformations, because the corresponding change of the state
vector does not amount to the appearance of a common phase factor.

If $x,y,z,t$ completely characterize the space-time, then the
$O(2\pi)$ and $R^2$ transformations coincide with the identical
transformation, which can change nothing physically. The
superselection rule is necessary in this case. If, in addition to
$x,y,z,t$, the spinor coordinates exist, $O(2\pi)$ and $R^2$ are
physically real transformations changing the sign of the
additional coordinates. In this case, the superselection rule is
not necessary. Thus, the proof of the possible existence of states
$\left| * \right>$ corresponding to the coherent superpositions of
states with even and odd numbers of fermions is simultaneously the
proof of the reality of a superspace with additional spinor
dimensions.

Such proof is presented below. Namely, we show that states $\left|
* \right>$ can be realized in a simple system with variable number
of electrons which form, together with an environment, an isolated
common system with a fixed number of electrons. In the case
considered, the interaction of the system with the environment can
be described as an external field acting on the system. This field
has the spinor character; i.e., it changes sign under the
$O(2\pi)$ and $R^2$ transformations. All eigenstates of the
Hamiltonian of the system are coherent superpositions $\left| *
\right>$ of the states with even and odd numbers of electrons, so
at temperatures well below the characteristic energy difference,
one (ground) of the states is automatically realized.

Irrespective of whether supersymmetry really exists or not, the
existence of additional dimensions of space-time results in the
existence of corresponding universal degrees of freedom for
systems of fermions. Below, the nature of these degrees of freedom
is elucidated for nonrelativistic systems. We show that they
correspond to the continuous change of the number of fermions in
the system. In principle, there are two ways to change the average
number $\left< N \right>$ of fermions continuously between two
neighboring integral values $N$ and $N+1$. The states with
nonintegral $\left< N \right>$ can be either impure states
corresponding to incoherent mixtures of states $N$ and $N+1$, or
pure states $\left| * \right>$ considered above. The first
possibility would be the only one if the superselection rule is
valid. We show that the change of $\left< N \right>$ in a system
from $N$ to $N+1$ through coherent superpositions $\left| *
\right>$ can be (and should be) interpreted as ``the motion'' of the
system ``as a whole'' along the additional fermionic dimensions of
space-time. Since the coherent change of $\left< N \right>$ is
accompanied by the change of the spin of the system, the
considered universal degrees of freedom include also the spin
degrees of freedom.

We consider condensed matter systems such as metallic
nanoparticles influenced by gate potentials at extremely low
temperatures. The nanoparticles behave like mesoscopic quantum
dots (MQD), i.e., all ordinary spatial degrees of freedom of the
particles are completely frozen out. Near some critical values of
the gate potentials, the particles possess electronic degrees of
freedom which are not still frozen and which are described
adequate in terms of the additional space-time coordinates.

We suggest experiments with MQDs which generalize the experiment
of Nakamura et al. \cite{n} on the observation of quantum
coherence between the states with different (but even in both
cases) numbers of electrons. The implementation of these
experiments will directly demonstrate the coherence between the
states with even and odd numbers of electrons and, thereby,
justify the reality of the new dimensions.

\section{Realization of states $\left| * \right>$}
We are going to demonstrate that the superselection rule is not
selfconsistent. We show that states $\left| * \right>$ can be
naturally realized in a simple realistic system with variable
number of electrons because this system is governed by the
Hamiltonian whose eigenstates are all coherent superpositions of
the states with even and odd numbers of electrons. The idea (see
 \cite{a3}) is as follows.

The number of electrons is a conserved quantity which is
analogous, in this respect, to the momentum. Physical systems with
Hamiltonians whose eigenstates are all coherent superpositions of
the states with different momenta are well known. A particle in an
external potential field depending on the particle coordinate is
the simplest example. In actuality, this particle is part of an
isolated system consisting of the particle and a certain massive
body, the interaction with which can be described as an external
field acting on the particle. As known, this is justified only if
certain conditions are fulfilled. First, the states of a massive
body must adiabatically adjust to the changes in the particle
coordinate in order to prevent excitation of the body degrees of
freedom. Second, all measurements must be made only with the
particle. No direct influence upon a massive body is possible.

Let us consider single electron with fixed spin projection (strong
external magnetic field is applied along $z$-axis) which can be
localized in one of two quantum dots (I and II) and can tunnel
from one dot to another. We are going to consider everything
inside the dot I as system with variable number of electrons we are
investigating (analogous to the particle in the above example) and
everything inside the dot II as environment (analogous to a massive
body).

This can be done by the following two steps. First, we consider
our initial system (single electron) as a part of a larger 
system with variable number of electrons consisting of both
quantum dots whose Hilbert space $S$ (see Fig.1) corresponds to
the general superposition \ket{} of all states
\ket{n,m} with the numbers of electrons $n$ in the dot I and $m$
in the dot II running through values $0,1$ independently. Let us
introduce in $S$ the operators $a, a^+$ and $b, b^+$ of
annihilation and creation of electrons in the dot I and the dot II,
respectively, so $n = a^+a$ and $m = b^+b$.

\begin{figure}[htb]
\begin{center}
\includegraphics[scale=0.5]{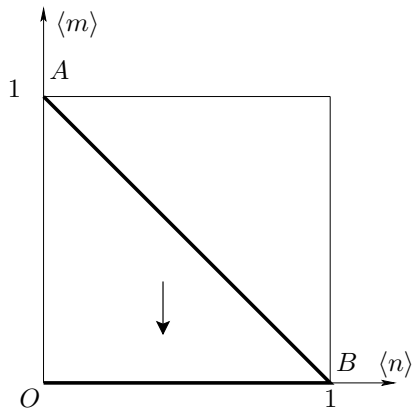}
\caption{Projection of the Hilbert space $s$ into the Hilbert space $s_I$}
\end{center}
\end{figure}

The Hilbert space $s$ of our initial system (the straight line
$AB$ in Fig.1) corresponds to the general superposition $\left|
 \right>_{i}$ of states $\left| 0,1 \right>$ and $\left| 1,0
\right>$. The characteristic feature of the space $s$ is that the
quantum number $m$ of the environment is determined unambiguously
by the quantum number $n$ of the system:

\begin{equation}
m \approx 1 - n. \label{eq2}
\end{equation}
Here the sign $\approx$ means, as in the book of Dirac \cite{dir}, that the
corresponding equality is the following auxiliary condition on the state vector

\begin{equation}
(m + n - 1)\left|\right>_i = 0, \label{eq3}
\end{equation}
but not
an operator identity in the whole space $S$. It is this property that allows
the interaction of the system with the environment to be treated as an external
field acting on the system. This is a fermionic analogy to the adiabatic
condition in the above example.

The second step is to project the Hilbert space $s$ of the system
into the Hilbert space $s_{I}$ (the straight line $OB$ in Fig.1)
of the dot I considered as a separate system. The corresponding state
vectors $\left| \right>_{I}$ are superpositions of states
$\left| 0 \right> \equiv \left| 0,0 \right>$ and $\left| 1 \right>
\equiv \left| 1,0 \right>$. The projection is the result of the
transformation $\left| \right> \rightarrow U \left| \right>$
of state vectors in the space $S$ with

\begin{equation}
U = n + \sigma (1 - n)b, \label{eq4}
\end{equation}
where $\sigma = \pm 1$.
We have

\begin{equation}
U \left| 0,1 \right> = \sigma \left| 0,0 \right> = \sigma \left| 0
\right>; \quad \quad
U\left| 1,0 \right> = \left| 1,0 \right> = \left| 1
\right>. \label{eq5}
\end{equation}

The operator $U$ is not unitary in $S$:

\begin{equation}
U^+U = n + (1 - n)m. \label{eq6}
\end{equation}
But in $s$, it satisfies the condition

\begin{equation}
U^+U \approx 1 \label{M}
\end{equation}
because

\begin{equation}
n + (1 - n)m \approx n + (1 - n)^2 = n + (1 - n) = 1. \label{eq8}
\end{equation}
The condition (\ref{M}) gives the possibility to introduce the transformed
Hermitian operator

\begin{equation}
F_I = UFU^+ \label{eq9}
\end{equation}
acting in $s_I$ for each Hermitian operator $F$ acting in $s$ in such a way
that the matrix elements do not change. In fact, let

\begin{equation}
\left| 1 \right>_I = U \left| 1 \right>_i, \quad \quad \left| 2 \right>_I = U
\left| 2 \right>_i \label{eq10}
\end{equation}
be two states in $s_I$ corresponding to two arbitrary states $\left| 1
\right>_i$ and $\left| 2 \right>_i$ in $s$. According to (\ref{eq10}),
(\ref{eq9}), and (\ref{M}) we have

\begin{equation}
\left< 2 \right|_I F_I \left| 1 \right>_I = \left< 2 \right|_i U^+UFU^+U
\left| 1 \right>_i = \left< 2 \right|_i F \left| 1 \right>_i. \label{eq11}
\end{equation}

The Hamiltonian of single electron can be represented in $s$ as

\begin{equation}
H = en + Em + Vab^+ + V^*ba^+, \label{eq12}
\end{equation}
where $e$ and $E$ are energies when electron is localized in the dot I
and the dot II, respectively, $V$ is the tunneling amplitude.

After simple calculations, we find quantities transformed to
$s_{I}$:

\begin{equation}
UnU^+ = n, \label{eq13}
\end{equation}

\begin{equation}
UmU^+ = nm + (1 - n)(1 - m) \approx 1 - n, \label{eq14}
\end{equation}

\begin{equation}
Uab^+U^+ = -\sigma a(1-m) \approx -\sigma a, \label{eq15}
\end{equation}
where the condition $m \approx 0$ in $s_{I}$ is used.

Finally, the Hamiltonian of the system interacting with the
environment is

\begin{equation}
H_{I} = UHU^+ \approx en + E(1 - n) + \eta a + \eta^*a^+, \label{eq16}
\end{equation}
where $\eta = -\sigma V$.

We note that $H_{I}$ does not contain operators $b$ and $b^+$ of
the environment. The interaction of the system with the
environment is described as an external field $\eta$ acting on the
system. The field $\eta$, as well as the operators $a$ and $a^+$,
and other spinor quantities, change sign under the $O(2\pi)$ and
$R^2$ transformations so that, for a given field value,
Hamiltonian (\ref{eq16}) is not invariant about these
transformations. Due to the presence of terms linear in electron
operators, all eigenstates of the Hamiltonian are coherent
superpositions of the states with even and odd numbers of
electrons. Actually, (\ref{eq16}) is the Hamiltonian of the two
level system
\begin{equation}
\left| * \right> = u\left| 0 \right> + v\left| 1 \right>.
\label{eq17}
\end{equation}

\section{Additional spinor dimensions}
Thus, to describe the above system correctly, we have to introduce
additional spinor dimensions of space-time. Assuming that the
corresponding coordinates are the nonrelativistic limit of the
coordinates considered in the field theory, we introduce (as in
\cite{a1},\cite{a2}) a Pauli spinor $\theta_{\alpha}$ where
$\alpha = 1,2$ is a spin index. Additional coordinates
$\theta_{\alpha}$ are Grassmann coordinates satisfying
anticommutation relations

\begin{equation}
\{\theta_{\alpha}, \theta_{\beta}\} = 0.
\label{A}
\end{equation}
Whatever the actual superspace structure is in the relativistic
case, this simplest possibility is quite general in the
nonrelativistic limit.

In the case considered above due to the presence of the strong
magnetic field along $z$-axis, the system (the dot I) should be treated
as ``moving'' along $\theta \equiv \theta_{1}$. Quantum mechanics
with anticommuting coordinates is well known (see for example \cite{lee}). The
``wave function'' $\Psi$ is an analitical function of $\theta$. Due to the
condition $\theta^2 = 0$, the most general $\Psi (\theta)$ is

\begin{equation}
\Psi (\theta) = uI +v\theta, \label{eq010}
\end{equation}
where $I$ is the unit of Grassmann algebra. By identifying $I =
\left| 0 \right>$ and $\theta = \left| 1 \right>$ we see that
states $\left| * \right>$ in (\ref{eq17}) are identical to
(\ref{eq010}). The physical change of states $\left| * \right>$
under the $O(2\pi)$ and $R^2$ transformations is connected with
the spinor nature of the physical coordinate $\theta$.

Operators $a^+$ and $a$ in the Hamiltonian (\ref{eq16}) play the
role of canonical coordinate $a^+ = \theta$ and momentum $a =
\partial/\partial \theta$ corresponding to the additional
dimension of space-time.

\section{Mesoscopic quantum dots}
To elucidate the physical meaning of the degrees of freedom corresponding to
the additional dimensions, it is very helpful to consider ground states of
a mesoscopic quantum dot at different values of the gate potential or at
different values of the electron chemical potential.

We consider a metal
particle with a large but finite number $N$ of electrons, connected by tunnel
junctions to macroscopic leads and (or) to other particles of the same type,
and being influenced by a gate potential. Let us suppose that the temperature
and all tunneling amplitudes are much smaller than the energy difference
$\delta\epsilon \sim E_{F}/N$ between the first excited and ground states of
the particle at a given number of electrons, $E_{F} \sim 10^{4}K$ is Fermi
energy. Under these conditions the metal particle behaves as a mesoscopic
quantum dot (MQD), i.e., it is a quasi-closed system in which all ordinary
degrees of freedom associated with spatial motion of electrons are completely
frozen. Inasmuch as even at lowest experimentally possible temperature $T \sim
1mK$ the number of electrons can not be larger than $10^7$, we have to deal
with metallic nanoparticles of the type obtained by Ralph et al.\cite{R}.

Thus, at a given $N$ the metal particle is in its ground state
$\left| N \right>$ with the energy $E_{N}$. Minimization of $E_{N}
= E_{N}(U) $ with respect to $N$ at a given gate potential $U$
determines the ground state value of $N = N(U)$. We see that in
MQD, a change in the number of electrons accompanying a change of
the gate potential occurs as a result of first-order phase
transitions between phases characterized by different integral
values of $N$ (see Fig.2). The jumps of the number of electrons occur at
critical values of the gate potential.

\begin{figure}[htb]
\begin{center}
\includegraphics[scale=0.5]{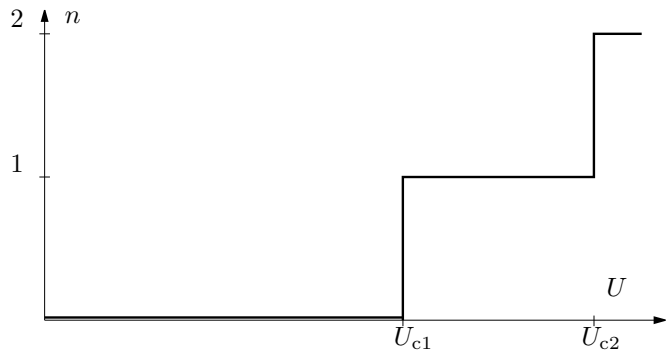}
\caption{Number of electrons $N=N_e+n$ vs gate potential $U$ at fixed $N_e$}
\end{center}
\end{figure}

It is important here to take into account the so-called parity
effect, i.e. the fact that in Fermi systems , the ground state
energy $E_{N}$ calculated with an accuracy appropriate for a
mesoscopic system contains in an explicit form the number of
particles in the combination $(-1)^N$. In order to deal only with
quasi-continuous functions at large $N$, we should introduce two
different functions $E_{N}^{o}$ and $E_{N}^{e}$, individually for odd and even
values of $N$, extrapolated to the same value of $N$. The difference $P = E_N^o
- E_N^e$, which is usually positive, can be considered as a quantitative
characteristics of the parity effect. Due to the parity effect, the steps
correspondong to even $N$ are longer than the steps corresponding to odd $N$.
(see Fig.2). With increasing $P$ the length of even steps increases, while
the length of odd steps decreases until the odd steps disappear at all. The
jumps of the number of electrons at critical values of the gate potential
become equal to 2. The phase transitions between states with neighboring even
numbers of electrons take place at large $P$. A clear example of a system
characterized by large $P$ is the so-called single-Cooper-pair box (see
\cite{n}), i.e. a superconducting MQD in which all electrons form Cooper-pairs
and the energy of an additional single electron (the superconducting energy
gap) is very large.

To classify all possible ground states of a {\it normal} MQD at different
values of the gate potential, let us put $N = N_e + n$ where the
quasicontinuous number $N_e \approx N$ ($N>>1$!) runs through all even numbers,
while $0 \leq n \leq 2$. There are three states for each $N_e$ with $n =
0,1,2$, respectively, the combinations ($N_e,n=2$) and ($N_e + 2,n=0$) being
identical.

Fig.2 shows the dependence $n = n(U)$ at a given $N_e$. There are two critical
values of the gate potential, $U_{c1} = U_{c1}(N_e)$ and $U_{c2} =
U_{c2}(N_e)$, corresponding to the conditions

\begin{equation}
E_{0}(N_e, U_{c1}) = E_{1}(N_e, U_{c1}),
\end{equation}
and

\begin{equation}
E_{1}(N_e, U_{c2}) = E_{2}(N_e, U_{c2}),
\end{equation}
respectively, where $E_{n}(N_e, U) \equiv E_{N_e + n}(U)$.

As the parity effect increases (with changing $N_e$), the quantities $U_{c1}$
and $U_{c2}$ approach one another, so that for certain $N_e = N_{ec}$ (or $P =
P_c$) determined by the equation

\begin{equation}
U_{c1}(N_{ec}) = U_{c2}(N_{ec}),
\end{equation}
a triple point may exist where all three states have the same energy. We note
that the triple point has actually been observed experimentally \cite{T}.

For an even larger $P$ ($P>P_c$) there is only one critical value,
$U_{c}(N_e)$, where

\begin{equation}
E_{0}(N_{e},U_c) = E_{2}(N_{e},U_c).
\end{equation}
The jump of $N$ at $U = U_c$ is equal to 2.

At critical values of the gate potential, the ground state is degenerate.
According to the quantum superposition principle (no superselection rule!),
there are infinite sets of ground states of the form

\begin{equation}
\left| * \right> = u \left| 0 \right> + v \left| 1 \right>,
\label{*}
\end{equation}

\begin{equation}
\left| * \right> = u \left| 1 \right> + v \left| 2 \right>,
\label{**}
\end{equation}
and

\begin{equation}
\left| e \right> = u \left| 0 \right> + v \left| 2 \right>,
\label{e}
\end{equation}
at $U = U_{c1}$ and $U = U_{c2}$ for $P < P_c$, and at $U = U_c$ for $P > P_c$,
respectively, where $\left| n \right> \equiv \left| N - N_e \right>$.

The ground states of the form (\ref{*}), (\ref{**}), and (\ref{e}) are
characterized by nonintegral averaged numbers of electrons $<N> = N_{e} +
|v|^2$, $N_{e} + 1 + |v|^2$, and $N_{e} + 2|v|^2$, respectively. One can say
that these states correspond to ``phase coexistence'' regions (vertical segments
in Fig.2) of first-order phase transitions taking place at $U = U_{c1}$,
$U_{c2}$, and $U_c$.

The ``phase coexistence'' of all three phases with $n = 0,1,2$,

\begin{equation}
\left| t \right> = (1 - w_{1} -w_{2})^{1/2} \left| 0 \right> + w_{1}^{1/2}
e^{i \phi_{1} } \left| 1 \right> + w_{2}^{1/2} e^{i \phi_{2} }
\left| 2 \right>
\label{t}
\end{equation}
takes place at the triple point $U =
U_{c1} = U_{c2}$. These ground states are characterized by two physical
(superconducting!, see \cite{a1,a2}) phases, $\phi_1$ and $\phi_2$, and $w_1
>0$, $w_2 >0$, $w_{1} + w_{2} < 1$.

A more general consideration is the following. Near critical values $U_{c1}$,
$U_{c2}$, and $U_c$ of the gate potential the MQD has some number of states
which are close in energy to the ground state. All other states have a much
higher energy and can be neglected in low-frequency dynamics. In this sense one
can say that the MQD possesses degrees of freedom active at low temperatures
and low frequencies (much below $\delta \epsilon$). Active degrees of freedom
are characterized by the Hilbert spaces (\ref{*}), (\ref{**}), (\ref{e}), or
(\ref{t}) near $U_{c1}$, $U_{c2}$, $U_c$, or near the triple point,
respectively.

To take spin into account, let us suppose that the state $\left| 0 \right>$
and the state $\left| 2 \right>$ are spin-singlets, and the state with $n = 1$
has a total spin 1/2. Otherwise these states are ``incompletely frozen'' with
respect to the Bose degrees of freedom. Then in the most general case
(realized near the triple point), four states, $\left| 0 \right>$, $\left| 1,
\alpha \right>$, and $\left| 2 \right>$ where $\alpha = 1,2$ is a spin index,
are close in energy. The Hilbert space of active degrees of freedom corresponds
to the general superposition

\begin{equation}
\left| g \right> = c_0 \left| 0 \right> + \Sigma_{\alpha} c_{1, \alpha} \left|
1, \alpha \right> + c_2 \left| 2 \right>,
\label{g}
\end{equation}
where $c_0$, $c_{1, \alpha}$, and $c_2$ are complex numbers. Active degrees of
freedom near $U_{c1}$, $U_{c2}$, and $U_c$, far from the triple point are
described by (\ref{g}) with $c_2 = 0$, $c_0 = 0$, and $c_{1, \alpha} = 0$,
respectively.

Let us show that the active degrees of freedom described by (\ref{g})
correspond to ``motion'' of the MQD along the spinor dimensions $\theta_{\alpha}$
of superspace. Due to the anticommutation relations (\ref{A}), the most general
wave function $\Psi (\theta_{\alpha})$ of the MQD is

\begin{equation}
\Psi (\theta_{\alpha}) = c_0 I + \Sigma_{\alpha} c_{1, \alpha} \theta_{\alpha}
+ c_2 \theta_1 \theta_2.
\label{S}
\end{equation}
By identifying $I = \left| 0 \right>$, $\theta_{\alpha} = \left| 1, \alpha
\right>$, and $\theta_1 \theta_2 = \left| 2 \right>$, we see that the
Hilbert space (\ref{g}) is identical to (\ref{S}).

The Hamiltonian of the MQD is expressed in terms of the coordinates
$\theta_{\alpha} \equiv a_{\alpha}^{+}$ and momenta $\partial/ \partial
\theta_{\alpha} = a_{\alpha}$ operators satisfying the canonical relations for
Fermi operators. The operators $a_{\alpha}$ and $a_{\alpha}^{+}$ represent
universal collective characteristics of any system of fermions under conditions
such that the Bose degrees of freedom are completely frozen. In exactly the
same manner, the conventional coordinate and momentum operators describe the
dynamics of a system with respect to a collective Bose degree of freedom under
conditions such that all other degrees of freedom are frozen.

The MQD is characterized by the following gauge invariant quantities
\begin{equation}
n = a_{\alpha}^{+} a_{\alpha}, \quad \quad {\bf S} = (1/2) a_{\alpha}^{+} 
{\bm \sigma}_{\alpha \beta} a_{\beta},
\label{O}
\end{equation}
where ${\bm \sigma}_{\alpha \beta}$ are the Pauli matrices. The operator $n$
corresponds to the parameter $n$ introduced above. The operator ${\bf S}$
is the operator of the total spin of the
MQD. The operators $a_{\alpha}$ and $a_{\alpha}^{+}$ are therefore a
generalization of the spin operators to the
case where, together with $x,y,z,t$, there exist additional dimensions described
by $\theta_{\alpha}$.

\section{Experiments}
The states (\ref{*}), (\ref{e}), and generally (\ref{g}) are the stationary
states of the MQD at critical values of the gate potential when the interaction
of the MQD with the environment or with other MQDs is neglected.
Manipulating by the gate potentials as functions of time in a system of two (or
more) MQDs one can realize conditions when even infinitesimal tunneling
amplitudes cause observable tunneling transitions of electrons between
different MQDs. This happens at the moments when the energies of different
states of the total system coincide, the states (\ref{*}), (\ref{e}), and
(\ref{g}) playing the role of asymptotic states. The coherence of these states
can be demonstrated by observations of interference phenomena (see \cite{n} and
below).

It is important to note that due to the relatively large dimensions of MQDs
one can measure the charge of a MQD without essential disturbance of
neighbouring MQDs. This can be done (as in the experiment of Nakamura et al.
\cite{n}) using a probe electrode connected to the MQD under study through a
tunneling contact or (as in the experiment of Aassime et al. \cite{D}) using a
probe electrometer based on a single-electron transistor.

Let us consider the simplest experimental situation. The role of quantum dots
(the dots I and II) in the discussion of Section II can be played by two MQDs.
The gate potentials should be close to corresponding critical values $U_{c1}$
for both MQDs. The sum of the parameters $n$ (determined by (\ref{O})) should be
equal to one. Under these conditions the tunneling of the additional (with
reference to the state with both $n$ equal to zero) single electron between
MQD I (considered as a system) and MQD II (considered as environment) is the
only active degree of freedom. All other degrees of freedom correspond to much
higher energies and can be neglected.

Thus, to prove experimentally the existence of the new dimensions of
space-time, one has to demonstrate the coherence of the superpositions
(\ref{eq17}) for the two level system (MQD I) governed by the Hamiltonian
(\ref{eq16}). The corresponding time-dependent Schroedinger equations are

\begin{equation}
i\dot{u} = \eta v , \quad \quad i\dot{v} = e(t) v + \eta^{*} u,
\label{sch}
\end{equation}
where the energy origin is chosen so that $E=0$. We suppose that the gate
potential of the MQD I can be varied to make the electron energy $e$ depending
on time: $e = e(t)$.

The system considered by Nakamura et al. \cite{n} to demonstrate the coherence
between states with different, but even in both cases, numbers of electrons is
also equivalent to a two level system described by the equations (\ref{sch}).
In the case considered by Nakamura et al. the role of the system (MQD I),
the environment (MQD II), and the spinor field $\eta$ are played by a
single-Cooper-pair box, a macroscopic superconductor (Cooper-pair reservoir),
and by the Josephson coupling constant, respectively.

Below we consider two experiments. The first experiment is the experiment of
Nakamura et al. performed with our two-level system (\ref{sch}). Before the
initial time ($t=0$), the two-level system has been in the ground state with
the gate potential of the MQD I such that $e>>|\eta|$. Accordingly, $u=1$ and
$v=0$. At $t=0$, the gate potential rapidly changes to a value for which
$e=0$. Then, the potential remains constant for a time $\Delta t$, after which
it rapidly regains its initial value. On the time interval between $t=0$ and
$t=\Delta t$, the system obeys Eqs.\ref{sch} with $e=0$ and initial conditions
$u(t=0)=1$ and $v(t=0)=0$. Then $u(t)=\cos |\eta|t$ and $v(t)=-i(\eta^{*}/|\eta|)
\sin |\eta|t$. At $t=\Delta t + 0$, one measures the everage charge of the
system

\begin{equation}
|v(\Delta t)|^2 = (1/2)(1 - \cos 2|\eta|\Delta t)
\label{st}
\end{equation}
as a function of pulse duration $\Delta t$. The observed oscillations
would indicate that the system coherently oscillates between the states with
$n=0$ and $n=1$ on the time interval ($0,\Delta t$). As pointed out in Section
III, the system considered is an oscillator moving along the fermionic
coordinate $\theta_1$.

The Nakamura-type experiment considered above can be modified by
passing from the single-pulse to two-pulse technique. As above, let the
two-level system be at $t<0$ in the ground state $u = 1$ and $v = 0$,
$e>>|\eta|$. The amplitude of the first gate-potential rectangular pulse is the
same as above (i.e., corresponds to $e=0$), but its duration is fixed at $t_1 =
\pi/4|\eta|$. Immediately after the pulse at $t=t_{1} +0$, the system is in
the state with $u=v=\sqrt{1/2}$. In the interval between $t=t_1$ and $t=t_{1}
+\Delta t$, the potential is equal to its initial value corresponding to
$e>>|\eta|$. Under this condition, the tunneling interaction of the system with
environment can be ignored and it behaves as a closed system in its pure state
characterized by

\begin{equation}
u(t)=\sqrt{1/2}, \quad \quad v(t)=-i(\eta^{*}/|\eta|)\sqrt{1/2}
\exp i\varphi(t),
\label{w}
\end{equation}
with the relative phase of the ground ($n=0$) and excited
($n=1$) states linearly depending on time: $\varphi(t) = -e(t-t_{1})$.

The second gate-potential pulse with parameters of the first pulse is switched
on at time $t_{1} + \Delta t$. Using Eqs.(\ref{sch}), one can see that, after
completion of the second pulse at time $2t_{1} + \Delta t$ ($\Delta t << t_1$
because $e>>|\eta|$), the everage charge of the system is

\begin{equation}
|v|^2 = 1/2(1 + \cos e\Delta t ).
\label{end}
\end{equation}
The observation of oscillations (\ref{end}) as a function of time delay $\Delta
t$ between the pulses would demonstrate that the state (\ref{w}) is realized.
This would be direct experimental proof of the reality of a superspace with
additional spinor dimensions.

\end{document}